


\input harvmac
\ifx\epsfbox\UnDeFiNeD\message{(NO epsf.tex, FIGURES WILL BE IGNORED)}
\def\figin#1{\vskip2in}
\else\message{(FIGURES WILL BE INCLUDED)}\def\figin#1{#1}\fi
\def\ifig#1#2#3{\xdef#1{fig.~\the\figno}
\goodbreak\midinsert\figin{\centerline{#3}}%
\smallskip\centerline{\vbox{\baselineskip12pt
\advance\hsize by -1truein\noindent\footnotefont{\bf Fig.~\the\figno:} #2}}
\bigskip\endinsert\global\advance\figno by1}

\def\ajou#1&#2(#3){\ \sl#1\bf#2\rm(19#3)}
\noblackbox
\def\a{\rightarrow}
\def\ap{\approx}
\def\d{\nabla}
\def\l{\lambda}
\def\R{Re^\phi}
\def\t{\tilde}
\def\({\left (}
\def\){\right )}
\def\[{\left [}
\def\]{\right ]}
\def\RN{Reissner-Nordstr\"om}
\font\ticp=cmcsc10
\gdef\journal#1, #2, #3, 19#4#5{
{\sl #1~}{\bf #2}, #3 (19#4#5)}

\lref\banks{T. Banks, A. Dabholkar, M. Douglas, and M. O'Loughlin,
\journal Phys. Rev., D45, 3607, 1992;
T. Banks and M. O'Loughlin, ``Classical and Quantum Production
of Cornucopions at Energies Below $10^{18}$ GeV,'' Rutgers preprint
RU-92-14, hep-th/9206055.}
\lref\wilczek{C. Holzhey and F. Wilczek,
\journal Nucl. Phys., B380, 447, 1992.}
\lref\brill{D. Brill,
\journal Phys. Rev., D46, 1560, 1992.}
\lref\nappi{M. McGuigan, C. Nappi, and S. Yost,
\journal Nucl. Phys., B375, 421, 1992.}
\lref\bizon{P. Bizon and T. Chmaj, ``Gravitating Skyrmions,'' Universitat Wien
UWThPh-1992-23, May, 1992.}
\lref\will{C. Will, {\it Theory and Experiment in Grvitational Physics},
Cambridge University Press (1981).}
\lref\dine{M. Dine, R. Rohm, N. Seiberg and E. Witten,
\journal Phys.
Lett., 156B, 55, 1986; L. Dixon, ``Supersymmetry Breaking in String
Theory,'' in the proceedings of the A.P.S. Division of Particle and
Fields Meeting Houston, TX 1990.}
\lref\witten{E.~Witten,
\journal Phys. Rev., D44, 314, 1991.}
\lref\ghs{D.~Garfinkle, G.~Horowitz and A.~Strominger,
\journal Phys.~Rev., D43, 3140, 1991; {\bf D45}, 3888({\bf E}) (1992).}
\lref\maeda{ G. Gibbons,
\journal Nucl. Phys., B207, 337, 1982;
G.~Gibbons and K.~Maeda,
\journal Nucl. Phys., B298, 741, 1988.}
\lref\sen{A. Sen, ``Rotating
Charged Black Hole Solution in Heterotic String Theory,'' Tata preprint
TIFR/TH/92-20, hep-th/9204046.}
\lref\rot{J.~Horne and G.~Horowitz,
\journal Phys.~Rev., D46, 1340, 1992.}
\lref\adm{R.~Arnowitt, S.~Deser and C.~W.~Misner,
\journal Ann.~of Physics, 33, 88, 1965.}
\lref\hawking{S.~Hawking,
\journal Phys. Rev. Lett., 26, 1344, 1971.}
\lref\strominger{For review of these puzzles and recent attempts to
resolve them, see J. Harvey and A. Strominger, ``Quantum Aspects of Black
Holes,'' EFI-92-41, hep-th/9209055.}
\lref\hindmarsh{A.~Hindmarsh, ``Odepack,'' in {\it Scientific Computing},
R.~Stepleman {\it et al.} eds. (North-Holland, Amsterdam, 1983), 55;
L.~Petzold,
\journal Siam J. Sci. Stat. Comput., 4, 136, 1983.
The software is publically available at {\bf netlib@research.att.com}.}
\lref\renata{R. Kallosh, A. Linde, T. Ortin, A. Peet, and A. Van Proeyen,
``Supersymmetry as a Cosmic Censor,'' Stanford preprint SU-ITP-92-13.}

\Title{\vbox{\baselineskip12pt\hbox{UCSBTH-92-17}
\hbox{YCTP-92-P37}
\hbox{hep-th/9210012}
}}
{\vbox{\centerline {Black Holes Coupled to a Massive Dilaton}
}}

\centerline{{\ticp James H. Horne}\footnote{$^\dagger$}
{Email address:
jhh@waldzell.physics.yale.edu}}
\vskip.1in
\centerline{\sl Department of Physics}
\centerline{\sl Yale University}
\centerline{\sl New Haven, CT 06511-8167}
\vskip .1in
\centerline{{\ticp Gary T. Horowitz}\footnote{$^*$}
{Email address:
gary@cosmic.physics.ucsb.edu}
}

\vskip.1in
\centerline{\sl Department of Physics}
\centerline{\sl University of California}
\centerline{\sl Santa Barbara, CA 93106-9530}

\bigskip
\centerline{\bf Abstract}

We investigate charged black holes coupled to a massive dilaton. It is
shown that black holes which are large compared to the Compton
wavelength of the dilaton resemble the \RN\ solution, while those
which are smaller than this scale resemble the massless dilaton
solutions.  Black holes of order the Compton wavelength of the dilaton
can have wormholes outside the event horizon in the string metric.
Unlike all previous black hole solutions, nearly extremal and extremal
black holes (of any size) repel each other. We argue that extremal
black holes are quantum mechanically unstable to decay into several
widely separated black holes.  We present analytic arguments and
extensive numerical results to support these conclusions.

\Date{10/92}

\newsec{Introduction}

There has been increasing interest in the properties of black holes in
string theory. One of the original motivations was the desire to gain
a better understanding of the allowed causal structure and possible
singularities in this theory. However, it has since been realized
that modifications arising from string theory may help to resolve some
of the puzzles associated with black hole evaporation~\strominger.
Although an exact black hole solution in string theory is now known in
two spacetime dimensions~\witten, we wish to consider the more
physical case of four dimensions where the exact black hole solution
is not known.  We can however work with the low energy string
equations of motion. Solutions to these equations should be close to
an exact string solution whenever the curvature is small compared to
the Planck scale.

Solutions to the low energy string equations of motion describing
black holes are known~\maeda\ghs, and differ from the standard \RN\
solution of general relativity when the charge is nonzero.  The reason
for this difference is the existence of a scalar field called the
dilaton.  Although the difference is negligible outside the event
horizon when the ratio of the charge to the mass is small, it becomes
significant near the extremal limit. In particular, in terms of the
metric that strings couple directly to, an extremal magnetically
charged black hole has neither a curvature singularity nor a horizon. In its
place is an infinite throat. The fact that a potentially infinite
volume of space may be present inside an object that appears small to
an observer outside the throat has been conjectured to play a role in
resolving the information problem that arises when black holes
evaporate~\banks.

Unfortunately, there is a serious omission in the investigations that
have been performed so far. In low energy string theory, the dilaton
is massless. But a truly massless dilaton violates the equivalence
principle and is inconsistent with observations~\will. Since the dilaton is
in the same supermultiplet as the graviton, it cannot get a mass while
maintaining supersymmetry.  However supersymmetry is certainly broken
at low energies, and it is widely believed (and hoped!)  that when
this occurs, the dilaton will become massive. Thus physically the most
interesting case is black holes coupled to a massive dilaton. This is
the subject of the present investigation.

We will show that for large black holes, the spacetime resembles the
\RN\ solution. When the charge is comparable to the mass, there is
both an outer (event) horizon and an inner (Cauchy) horizon. In the
extremal limit, these two horizons coincide.  For small black holes,
the solution resembles the massless dilaton case. There is only a
single horizon, and in the extremal limit, it moves off to infinity
leaving an infinite throat.  So infinite throats exist even in the more
realistic theory
with a massive dilaton.  The scale at which the black hole
changes character is simply the inverse mass, or Compton wavelength,
of the dilaton.  This is physically reasonable.  When the
Schwarzschild radius is less than the Compton wavelength, the black
hole does not see the mass, and behaves like the massless case.  When
the Schwarzschild radius is much larger, the dilaton is essentially
fixed at its minimum value and does not change enough to modify the
geometry. Numerically, if the dilaton has a mass of 1 TeV, the
transition occurs at a black hole mass of approximately $10^{11}$ gms.
So the dilaton does not affect solar mass sized black holes, but could
play an important role in the evaporation of smaller holes.

Charged black holes with a massless dilaton (and $\phi \a 0$ at infinity)
only exist when $Q^2 \le
2M^2$.  This upper limit on the charge for a given mass is larger than
the usual bound $Q^2 = M^2$ without the dilaton. We will see that
black holes with a massive dilaton have a maximal charge to mass ratio
which depends on the size of the black hole. It appears to vary continuously
from $Q^2 / M^2 =2$ for small black holes to $Q^2 / M^2=1$ for large
ones. The fact that the charge can be greater than the mass has the
following important consequence.  Two widely separated \RN\ black
holes will attract each other unless $Q=M$. Since this is the extremal
limit, it is not possible for black holes to be repulsive. When a
massless dilaton is added, it adds an extra attractive force. So even
though the extremal limit has a larger charge to mass ratio, this
limit again corresponds to the point where the forces cancel at large
separations.  With a massive dilaton, the situation is different. We
will see that at large distances, the solution always approaches the
\RN\ metric, and the force associated with the dilaton is negligible.
Since the presence of the dilaton near the horizon allows $Q^2 > M^2$,
one has black holes which (at large separations) repel each other!  To
the best of our knowledge, these are the first examples of
gravitationally bound objects which are repulsive.  Since $Q=M$ is not
the extremal limit, initially static black holes will Hawking radiate
and start to repel each other. We will argue that one consequence of
this repulsive force is that large extremal black holes are quantum
mechanically unstable to decay into multiple extremal black holes.

Another unusual property of black holes coupled to a massive dilaton
is the presence of wormholes.\foot{We will use the word ``wormhole''
to describe a region of space where the area of spheres decreases and
then increases again.
These are three dimensional -- not four dimensional
-- objects. See~\adm\ for another example of this type of wormhole.}
It is well known that the maximally extended Schwarzschild solution
contains a wormhole in the sense that a spacelike surface connecting
the two asymptotically flat regions reaches a minimum size inside (or
on) the event horizon. This wormhole, however, cannot be transversed
by physical observers since it quickly collapses to zero size.  We
will see that the string metrics describing certain black holes with a
massive dilaton have wormholes {\it outside} the event horizon. These
wormholes are static and can be transversed by strings falling into
the black holes.

The exact form of the dilaton potential in string theory is not known.
We will consider the simplest choice $m^2 \phi^2$ and later briefly
discuss what effect more realistic choices would have. Of course, for
a general potential, when $\phi$ is small and near its minimum, $m^2
\phi^2$ is a good approximation.  Adding this term to the standard low
energy string action yields
\eqn\lowenergy{  S=\int d^4x \left[{\cal R} - 2(\d\phi)^2  -2m^2\phi^2
                 - e^{-2\phi} F^2 \right],}
where $\cal R$ is the scalar curvature and $F_{\mu\nu}$ is the Maxwell
field strength.  As indicated in \lowenergy, we will work with the
metric having the standard Einstein action. This is related to the
metric that the strings couple directly to via the conformal rescaling
$\hat g_{\mu\nu} = e^{2 \phi} g_{\mu\nu}$.
Both metrics have the same causal structure except in the case of
extremal magnetically charged black holes, where a singular horizon in
the Einstein metric corresponds to an infinite throat in the string
metric.  The equations of motion of \lowenergy\ are
\eqna\eom
$$\eqalignno{   0 & =   \d_\mu (e^{-2\phi}F^{\mu\nu}) \>, & \eom a \cr
     0 & = \d^2 {\phi}- m^2 \phi + \half e^{-2\phi}F^2 \>,& \eom b \cr
       {\cal R}_{\mu\nu} & = 2 \d_\mu \phi  \d_\nu \phi +
		      2 e^{-2\phi} F_{\mu\rho} {F_\nu}^\rho
		    + g_{\mu\nu} [ m^2 \phi^2 - \half e^{-2\phi} F^2]
           \>.  & \eom c \cr} $$
These equations are invariant under the same electromagnetic duality
transformation that holds when $m=0$. Namely, one can replace
$F_{\mu\nu}$ and $\phi$ with $\t F_{\mu\nu} \equiv \half e^{-2\phi}
\epsilon_{\mu\nu}^{~~~\lambda\rho}F_{\lambda\rho}$ and $\t \phi = -\phi$.
We will consider a purely magnetically charged black hole: $ F = Q
\sin \theta\ d \theta \wedge d\varphi $. Electrically charged holes
can be obtained by applying the above duality.

We are interested in static, spherically symmetric solutions.
Unfortunately, we do not expect that exact solutions can be expressed
in a simple closed form. We will study these solutions using two
different forms of the metric. We will first choose coordinates so
that the metric takes the form
\eqn\einone{ds^2 = -\l dt^2 + {dr^2 \over \l} + R^2 d\Omega \>,}
where $\l$ and $R$ are functions of $r$ only.  This has the advantage
that both the massless dilaton black holes and no dilaton black holes
(i.e. the \RN\ solutions) are simply expressed in these coordinates.
This will facilitate a discussion of approximate solutions as well as
lead to some general properties. In Sec.~2 we will discuss properties
common to all black hole solutions, while in Sec.~3 we will consider
separately black holes of different sizes.  In Sec.~4 we will present
our numerical results. For this purpose, it is more convenient to
choose the following form of the metric
\eqn\eintwo{ ds^2 = -fdt^2 + {dr^2\over h} + r^2 d\Omega \>,}
where $f$ and $h$ are functions of $r$ only. With this form of the metric,
$f$ can be solved for explicitly, and the $h$ equation becomes first
order instead of second order. Sec.~5 contains some concluding remarks.

\newsec{General Properties}

Choosing the form of the metric~\einone, the dilaton equation \eom{b} and
three nontrivial components of the metric equation \eom{c} become
\eqna\eineom
$$ \eqalignno{
  (R^2 \l \phi')' & = m^2 R^2 \phi - {Q^2 e^{-2\phi} \over R^2}  \>,&
                                                         \eineom a \cr
(R^2 \l')' & = -2m^2 R^2 \phi^2 +
         { 2 Q^2 e^{-2\phi} \over R^2} \>, & \eineom b \cr
(\l (R^2)')' & = 2 -2m^2 R^2 \phi^2 -{2Q^2 e^{-2\phi} \over R^2} \>,
                                      & \eineom c \cr
 0 & = R'' +R(\phi')^2  \>. & \eineom d \cr }$$
These four equations are not all independent. Eqs.~\eineom{c}
and~\eineom{d} can be combined to yield a single equation with no
second derivatives.

When $m=0$ these equations have a simple exact black hole solution~\maeda\ghs
\eqn\massless{
\l= \( 1- {2M\over r}\),\;\;\;
  R^2 = r^2 \(1-{Q^2 e^{-2 \phi_{\infty}} \over Mr}\) ,\;\;\;
   e^{-2\phi} = e^{-2 \phi_{\infty}} \(1 -
                      {Q^2 e^{-2 \phi_{\infty}} \over Mr}\) ,}
where $M$ is the mass and $\phi_{\infty}$ is the value of the dilaton at
infinity.  The metric is identical to Schwarzschild except that the
area of the spheres of spherical symmetry are reduced by an amount
depending on the charge.  This area goes to zero at $r = {Q^2 e^{-2
\phi_{\infty}} \over M}$ which is the curvature singularity. In the extremal
limit, $Q^2 = 2M^2 e^{2 \phi_{\infty}}$, the horizon itself becomes
singular.  However, when we conformally rescale to the string metric,
one finds that the extremal limit is given by
\eqn\extremal{ \widehat{ds}^2 = -dt^2 +
\( 1-{\sqrt 2 |Q|\over r} \)^{-2}dr^2 + r^2 d\Omega}
where we have absorbed the constant $e^{\phi_{\infty}}$ into a
redefinition of $r$ and $t$. This metric has neither a horizon nor a
curvature singularity. The surfaces of constant time are
asymptotically flat, and approach an infinite cylinder as $r$
approaches $\sqrt 2 |Q|$.

We begin our discussion of black holes with $m \ne 0$ by deriving a
few general properties of these solutions.  As boundary conditions at
infinity, we require that the metric be asymptotically flat and the
dilaton vanish.  (Since we have chosen the minimum of the dilaton
potential to be at $\phi=0$, one cannot shift the dilaton at infinity
without destroying asymptotic flatness, unlike the massless case.) In
particular, at infinity, $R$ must be equal to $r$. So integrating
\eineom{d} from any $r$ to infinity, we get $R'(r) = 1 +\int_r^\infty
(\phi')^2 R\,d\tilde r $.  In particular, $R'\ge 1$ everywhere and $R$
is monotonically increasing.  This shows that there are no wormholes
outside the horizon in the Einstein metric.

Now consider the behavior of the dilaton outside the horizon.  We
first show that the dilaton must be positive on the horizon.
Multiplying the dilaton equation by $\phi$ yields
\eqn\dileq{ (\phi R^2 \l \phi')' = R^2 \l{\phi'}^2 + m^2 R^2 \phi^2
-\phi Q^2e^{-2\phi}/R^2 \>.}
If $\phi$ were negative on the horizon one could integrate \dileq\
from the horizon out to $\phi = 0$. The left hand side vanishes and
the right hand side is positive definite, yielding a contradiction.
Next, it follows immediately from the dilaton equation that at a
negative extremum, $R^2 \l \phi'' < 0 $. This means that if the
dilaton ever becomes negative, it must continue to decrease.  It
cannot have a local minimum with $\phi < 0$. Since the dilaton is
positive at the horizon and zero at infinity, it must remain
nonnegative.  Finally, one can show that the dilaton is monotonically
decreasing outside the horizon.\foot{We thank Ruth Gregory for
suggesting this argument.} Suppose $\phi$ has a local maximum. Then
the left hand side of eq.~\eineom{a}\ must be negative. This implies that
\eqn\philaw{  m^2\phi e^{2\phi} < Q^2/R^4 \>.}
Since $\phi$ is decreasing as one moves in, the left hand side is
decreasing while the right hand side is increasing.  So the inequality
remains valid everywhere between the local maximum and
the horizon.  ($\phi$ cannot have a
local minimum since that would require that the inequality be
reversed.)  This leads to a contradiction at the horizon,
since the left hand side of \eineom{a}\ is
$R^2 \l' \phi'$ at the horizon, which is positive if $\phi' >0.$
This violates the
inequality~\philaw. Since $\phi$ cannot have a local maximum,
it must be monotonically decreasing outside the horizon.

Next, consider the solution in the region far from the black hole.
Since the metric is asymptotically flat and $\phi$ vanishes at
infinity, the dilaton equation
\eineom{a} reduces to
\eqn\dilasym{ (r^2 \phi')' - m^2 r^2 \phi = - Q^2/r^2 . }
For large $r$ the derivative term is negligible and the solution is
\eqn\dilbsym{ \phi \approx  {Q^2 \over m^2 r^4} \>.}
The massless dilaton falls off at infinity like $1/r$, and a massive
scalar field with localized sources falls off exponentially.  Here we
have both a mass and a source which is going to zero like a power of
$1/r$. These combine to give the somewhat unusual asymptotic behavior
of the dilaton. Now consider the metric equations~\eineom{b}
and~\eineom{c}.  The contributions to the right hand sides from the
dilaton fall off like $1/r^6$ which is much faster than the dilaton
independent term $Q^2/r^2$.  As a result, the leading order solution
for $\l$ and $R$ is exactly the \RN\ solution.  The first order
corrections to \RN\ can be found by solving the equations
perturbatively. One finds that
\eqna\asym
$$\eqalignno{ \l & \ap 1- {2M\over r} +{Q^2 \over r^2} -
         {Q^4 \over 5m^2 r^6} \>,  & \asym a\cr
   R & \ap r \(1 -{2Q^4\over 7m^4 r^8}\) \>. & \asym b\cr }$$
Thus the corrections to the \RN\ solution are quite small for large
$r$. We should caution the reader at this point that the full power
series expansion of the solution around $r = \infty$ has zero radius
of convergence, so one must be careful when including large numbers of
higher order terms in~\asym{}. The approximations we have written down
give the correct leading behavior.

Now we consider the behavior of the solutions near the singularity.
Consider first the massless dilaton solution \massless.  Since the
dilaton is diverging near the singularity at $r = {Q^2 e^{-2 \phi_{\infty}}
\over M}$, it might appear that the effect of adding a nonzero mass
will become important.  However the dilaton only diverges
logarithmically, and in \eineom{} it appears multiplied by $R^2$
which vanishes linearly.
(By contrast, the charge terms remain nonzero near the singularity.)
The net effect is that the mass term becomes {\it less} important as
one approaches the singularity. So the $m=0$ solution \massless\
becomes a better and better approximation to an $m \ne 0$ solution in
this limit.  However this is not sufficient to prove that the massive
dilaton black holes approach the $m=0$ solution \massless\ near the
singularity.  One knows only that there are some solutions of the
$m\ne 0 $ equations which agree with \massless\ initially
and start to deviate as one moves out. There is no guarantee that the
resulting solutions will be asymptotically flat, or even have an event
horizon.

We will see in Sec.~4 that the massive dilaton black holes do approach
a solution of the $m=0$ equations near the singularity, but not
necessarily the one given in eq.~\massless.  The solution \massless\
was determined by the requirement that it have a regular event horizon
and be asymptotically flat. Near the singularity, these conditions do
not apply. The general solution to equations~\eineom{} with $m=0$ can
be found in closed form.
This solution
depends on five free parameters, but it is convenient to express it in terms
of six parameters subject to one condition. Let
$x= \({r-r_+\over r-r_- }\)^{1\over 2}$.
Then the solution is given by
\eqn\general{\l ={{x^{\sqrt{2-c_1^2}}} \over c_2 x^{c_1} + c_3 x^{-c_1}}
\>,\;\;\;\;\;
R^2 = (r-r_+)(r-r_-)/\l
\>,\;\;\;\;\; e^{2(\phi -\phi_{\infty})} = x^{2\sqrt{2-c_1^2}}/\l \>,}
where $c_1^2 c_2 c_3 (r_+ - r_-)^2 + 2Q^2 e^{- 2 \phi_{\infty}}=0$.
The five free parameters are just what one expects for the general
solution to three second order differential equations and one
constraint. However two of these parameters are essentially pure
gauge. The form of the metric~\einone\ is invariant under shifting $r$
by a constant, and rescaling $r\a cr$, $\l \a c^2 \l$ and $t \a t/c$.
So there is really a three dimensional space of physically different
solutions. When $c_1 = 1$ the solution~\general\ reduces to the
standard massless dilaton black hole, with the two remaining
parameters corresponding to $M$ and $\phi_{\infty}$. When $c_1 \ne 1$
the solution \general\ does not have a regular event horizon.  One can
verify that unless $c_1$ is close to $\sqrt{2}$, the effect of a mass
term becomes negligible near the singularity for any of these
solutions.

\newsec{Approximate Solutions}

There are three distinct types of black hole solutions depending on the
size of the black hole compared to the Compton wavelength
of the dilaton.

\subsec{Large Black Holes}

First consider a black hole with $Mm \gg 1$ i.e.~a black hole with
Schwarzschild radius much larger than the Compton wavelength of the
dilaton.  We claim that the asymptotic solution~\dilbsym, \asym{} found above
remains close to the exact solution up to and past the event horizon.
Outside the horizon, the maximum possible value of the correction
terms to \RN\ is obtained when $Q^2 \ap M^2$ and $r\ap M$. Even in this
case, the correction to $\l$ is of order $(Mm)^{-2}$ and the
correction to $R$ is of order $(Mm)^{-4}$. Since these correction
terms were derived under the assumption that $\phi \ap Q^2/m^2 r^4$ we
must also verify that this remains valid everywhere outside the
horizon. This solution for $\phi$ followed from the fact that $\phi
\ll 1$ and the derivative term in \dilasym\ was negligible. The first
condition is clearly satisfied until one is well inside the black
hole.  The second condition will remain true until
\eqn\rnbad{ [(r^2 -2Mr +Q^2) \phi']' \ap Q^2 /r^2 .}
Using the approximate value of $\phi$ from \dilbsym\ one finds that
the approximation is valid as long as $m r > (m M)^{1/3}$ or $m r > (m
Q)^{1/2}$ whichever is larger.  Since the event horizon occurs when
$r$ is of order $M$, this condition is also valid until well inside
the black hole.  (The second alternative is just the condition that
$\phi$ be of order one.)  For $Q^2$ of order $M^2$, the inner horizon of
the \RN\ black hole
is close to the event horizon. The above argument indicates that
the solution will stay close to \RN\ even past the inner horizon. This
suggests that  large black holes with large charge
coupled to a massive dilaton will have
nonsingular inner horizons even though massless dilaton black holes do
not.\foot{{\it Rotating} black holes with a massless dilaton do have an
inner horizon~\sen.}  The numerical solutions in Section 4 will
confirm this. If the charge is much less than the mass, the approximation
breaks down before the \RN\ inner horizon is reached. In this case, we will
find numerically that the solutions have only one horizon.

For a \RN\ black hole, the extremal limit occurs when the inner and
outer horizons coincide. This happens when $Q^2 = M^2$ and the horizon is
$r_0
 = M$.  Since the large black hole is quite similar to  \RN\
 until well within the inner horizon for $Q^2 \sim M^2$, the
extremal limit of the large black hole is qualitatively similar to the
\RN\ extremal limit.  The presence of a massive dilaton does however
shift the extremal limit away from $Q^2 = M^2$.  With the
parameterization in \einone, a horizon occurs whenever $\l = 0$, so
using~\asym{}, the condition for a horizon is
\eqn\horcon{ r^2 -2Mr +Q^2 = {Q^4 \over 5m^2 r^4}\>. }
We can calculate the deviations from the \RN\ limit by
fixing $M$ and perturbing $Q^2 \rightarrow M^2 + \delta Q^2$ and
$r_0 \rightarrow M + \delta r$. To leading order in $1/(mM)$,
\horcon\ gives  $\delta Q^2 = 1/5m^2$, or
\eqn\qmod{  {Q^2_{\rm ext} \over M^2} = 1+ {1\over 5M^2 m^2}\>.}
Thus in the extremal limit, the charge is always larger than the mass.
The maximum charge for given mass increases as the black hole becomes
smaller. If the charge is increased beyond the extremal limit, the
horizons disappear and a naked singularity occurs.

The electromagnetic and gravitational forces between two extremal \RN\
black holes (of the same sign) exactly cancel. We have seen that large
extremal string black holes approach \RN\ asymptotically and always
have $Q^2_{\rm ext} > M^2$.  This increases the magnetic repulsion, so
two extremal string black holes at large separation will repel each
other. This fact has the following important consequence. It is
energetically favorable for one large extremal black hole to split
into several smaller ones.  This follows from~\qmod\ since the mass of
a single extremal black hole with charge $Q$ is $M_1 = (Q^2 -
1/5m^2)^{1\over 2}$ while the mass for $n$ widely separated extremal
black holes with charge $Q/n$ is
\eqn\separate{ M_n = n\[\({Q\over n}\)^2 - {1\over 5m^2} \]^{1\over 2}
 = \[Q^2 - {n^2\over 5m^2} \]^{1\over 2}  \>. }
Clearly, $M_n$ is a decreasing function of $n$. (We cannot decrease
$M_n$ to zero since this formula is only valid when $mM_n \gg 1$.)  Of
course this breakup cannot occur classically since horizons cannot
bifurcate~\hawking. But it will occur through quantum tunneling, if
there is no infinite barrier to prevent it. One can argue against the
presence of an infinite barrier by finding an instanton which
describes this tunneling event.  This is currently under
investigation.

The possibility that extremal \RN\  or massless dilaton black holes can quantum
mechanically bifurcate has recently been discussed~\brill\renata.  Since there
is no force between these black holes, the analogous calculation would
show $M_n$ is independent of $n$. The different states are degenerate
in energy. Whether or not tunneling is possible in this case, we have
shown that it is much more likely to occur when there is a massive
dilaton.

Although we have considered extremal black holes, it should be clear
that the same conclusion applies to nearly extremal black holes. The
fact that $M_n < M_1$ will not be altered if the charge on each black
hole is decreased slightly. The main difference between extremal and
nonextremal black holes, is that the nonextremal holes will quantum
mechanically emit ordinary Hawking radiation.  Since the extremal
black holes have a degenerate horizon, their Hawking temperature
vanishes. As we have just seen, this is not sufficient to conclude
that they are quantum mechanically stable. This strongly suggests that
the extremal black holes are not supersymmetric.

\subsec{Small Black Holes}

Now consider the case $mM \ll 1$ when the black hole is small compared
to the Compton wavelength of the dilaton. In the asymptotically flat
region far from the hole, the dilaton satisfies eq.~\dilasym\ and has
the asymptotic solution \dilbsym.  Since $mM \ll 1$, there is now a
region $(mM)^{1/3} \ll mr \ll 1$ where the metric is still essentially
flat, but the dilaton is no longer given by \dilbsym. (The lower limit
comes from requiring that the terms in the dilaton equation depending
on $M$ be negligible.) The exact solution to \dilasym\ can be found in
terms of the exponential integral function, but for our purposes it
suffices to notice that for $mr \ll 1$, the solution is approximately
\eqn\phiapp{ \phi \approx -{Q^2 \over 2r^2} + {A\over r} }
where $A$ is an arbitrary constant. We showed earlier that the dilaton
cannot be negative outside the event horizon. Thus, in order to have a
solution describing a black hole, $A$ must be sufficiently large that
the second term dominates the first.  But this is precisely the
behavior of a massless dilaton at large distances from a black hole.
One expects that the solution will now resemble the massless dilaton
black hole, at least until the dilaton becomes large.

When considering how the massless dilaton black hole matches onto the
\RN\ solution, it is perhaps worth mentioning that the two solutions
are actually more similar than they may appear. As we have mentioned,
there is a residual
gauge freedom in the form of the metric \einone\ corresponding to
shifting $r$ by a constant.  If we introduce a new radial coordinate
$r = \t r + Q^2/2M$, then the massless dilaton solution \massless\
(with $\phi_{\infty} = 0$) becomes
\eqn\newrad{ \l \ap 1 - {2M \over \t r } + {Q^2 \over \t r^2} +O(\t r^{-3})
 \qquad R^2 \ap \t r^2 \[ 1- \({Q^2 \over 2M\t r} \)^2 \] +O(\t r^{-1})}
which shows that the solutions really differ only at order $1/r^2$.

In the general massless dilaton solution~\general, a horizon exists
only when $c_1 = 1$. Since we expect that small black holes resemble
the massless dilaton solution, $c_1$ must be 1 up to small
corrections.  Small deviations from $c_1=1$ are allowed since the corrections
from the dilaton mass can compensate and still permit a horizon to form.
We will see in the next section that for nonextremal black holes, these small
deviations from $c_1=1$ do not vanish as one approaches the singularity. Thus,
even though the mass terms become negligible near the singularity, the
solution is not approaching~\massless. However, in the extremal limit,
the horizon itself approaches the singularity. So the small corrections to
$c_1$ must be getting smaller and smaller. Thus, up to the coordinate freedom
of shifting $r$
and rescaling, the solution approaches the standard massless
solution~\massless. In the string metric, the extremal small black
holes will disappear down an infinite throat.
One might be tempted to conclude that the extremal value of the
charge for any small black hole is given by the massless dilaton result
$Q^2_{\rm ext} = 2M^2 $ (using the fact that the dilaton must vanish at
infinity). However, the
parameters $M$ and $\phi_{\infty}$ in~\massless\  do not in general
correspond exactly to the asymptotic values of the mass or
dilaton. This should be the case only
in the limit
of very small black holes. Thus we expect
that the extremal value of the charge for small
black holes has the form
\eqn\qmodb{ {Q^2_{\rm ext} \over M^2} =
                        2 - \alpha m^2 M^2 + {\cal O}(m^4 M^4)}
where $\alpha$ is an undetermined positive constant. It is simple to
check that extremal black holes with charge given by~\qmodb\ are
unstable in exactly the same manner as the large black holes.

\subsec{Black Holes with $Mm \ap 1$ }

Some analytic results about black holes with $Mm \sim 1$ can be
obtained as follows. We have seen that the extremal limit of a large
black hole has a degenerate horizon, while the extremal limit of small
black holes has a singular horizon (which in the string metric
corresponds to an infinite throat). Let us consider the condition for
the existence of a degenerate horizon. If both $\l$ and $\l'$ vanish
at a point $r_d$, then from \eineom{a} and \eineom{c},
\eqn\dfirst{ \eqalign{
 m^2 R_d^2 \phi_d & = {Q^2 e^{-2\phi_d}\over R_d^2} \>,\cr
 m^2 R_d^2 \phi_d^2 & + {Q^2 e^{-2\phi_d}\over R_d^2} = 1 \>,}}
where $R_d$ and $\phi_d$ are the values of $R$ and $\phi$ at $r_d$.
Combining these equations yields
\eqn\rcon{ R^2_d = {1 \over m^2 \phi_d (1 + \phi_d)}\>,}
\eqn\dcon{  {e^{2\phi_d} \over \phi_d (1+ \phi_d)^2} = m^2 Q^2 \>.}
Eq.~\dcon\ has two different regimes depending on the value of $m^2
Q^2$. The left hand side reaches its minimum value of $e^2/4$ when
$\phi_d = 1$.  Thus degenerate horizons cannot occur when $m^2 Q^2 <
e^2/4$.  In other words, black holes with $m^2 Q^2 < e^2/4$ cannot
have an extremal limit similar to the \RN\ solution. Instead, they will behave
like the massless dilaton solution.  When $m^2 Q^2$ is larger than the
minimum value, there are two solutions for $\phi_d$.  They can be
distinguished by looking at $\l''_d$ which from \eineom{b} is
\eqn\tcon{ R_d^2 \l_d'' = 2m^2 R_d^2 \phi_d(1-\phi_d) \>.}
Since $\l > 0$ outside the degenerate horizon (assuming it is the
event horizon), we require $\l''_d>0$ which implies $\phi_d < 1$.  For
a large black hole with $m^2 Q^2 \gg 1$, the solution of \dcon\ for
$\phi_d$ is approximately $\phi_d = (mQ)^{-2}$. This agrees with our
earlier solution $\phi = Q^2/m^2 r^4$ when we use the fact that $r =
Q$ at the horizon of degenerate large black holes.

The critical value $m^2 Q^2 = e^2/4$ is quite interesting.  We see
from \tcon\ that $\l''_d = 0$ when $\phi_d=1$, so that the degenerate
horizon is actually {\it triply} degenerate at the critical value (the
higher derivatives are nonzero). This may indicate that three horizons
are coming together at this point instead of just the two that come
together when $m^2 Q^2 > e^2/4$. The possibility of black holes with
three horizons has been discussed in the literature \nappi.  However,
we see no evidence of this in the numerical simulations described in
the next section. Because of the singularities in the equations, the
numerical simulations behave very poorly right near this critical
value, so we cannot rule out spacetimes with three horizons. However,
it is clear that this is not the generic case for $m^2 Q^2 > e^2/4$.

All of the preceding discussion has been in term of the Einstein
metric.  Strings couple instead to the string metric, which is equal
to the Einstein metric multiplied by $e^{2\phi}$. So the radius of the
spheres of spherical symmetry is $Re^\phi$. We have shown that $R$ is
monotonically increasing and $\phi$ is monotonically
decreasing outside the horizon.
So $(\R)'$ need not have a definite sign.  For the massless dilaton
black holes one finds that $\R$ is an increasing function
of $r$, which means that the spacetime is similar to the Schwarzschild
solution.  The maximally extended spacetime has a wormhole at the
horizon, but an observer falling in sees the spheres monotonically
decrease in area. When $m\ne 0$, the situation is different: $Re^\phi$
can decrease outside the horizon of certain black holes. Since $\R$ is
certainly increasing near infinity, the spacetime describes a wormhole
{\it outside} the event horizon.

To see this, we will need a relation between $R'$ and $\phi'$ valid
for any degenerate horizon. Taking the derivative of the dilaton
equation of motion and evaluating at a degenerate horizon yields
\eqn\worms{    R'_d = -R_d \phi'_d {1+2\phi_d^2 \over 4\phi_d} \>.}
Since $R' >1$ everywhere, we see that $\phi$ must decrease away from a
degenerate horizon as well as a regular horizon.  Using \worms\ we
obtain:
\eqn\wormy{   (\R)'_d = e^{\phi_d}( R' + R\phi')_d
	     =R_d\phi'_d e^{\phi_d} \[1- {1+2\phi_d^2 \over 4\phi_d} \] \>.}
Since $\phi'_d <0$, if $\phi_d$ is either very large or very small the
quantity in brackets is negative and $(\R)'_d>0$ as expected. However
for $1-1/\sqrt 2 < \phi_d < 1+1/\sqrt 2 $ one has $(\R)'_d<0 $. In this
case the string metric describes a wormhole outside the event horizon.
A string falling in will pass through the throat of the wormhole
before entering the black hole.

The above local argument is not complete in that we have not yet shown
that one can evolve the spacetime outward from the wormhole and obtain
an asymptotically flat region. This gap will be filled by the
numerical solutions in the next section. Although we have found these
wormholes by studying degenerate horizons, it should be clear that
wormholes can exist outside regular event horizons as well.  If one
increases the mass slightly keeping the charge fixed, the throat of
the wormhole will remain outside the horizon.

\newsec{Numerical Results}

\subsec{General Method}

As mentioned earlier, for numerical results, it is convenient to use
the following form of the Einstein metric
\eqn\einthree{   ds^2 = -fdt^2 + h^{-1}dr^2 + r^2 d\Omega }
where $f$ and $h$ are functions of $r$ only. This form of the metric
is justified since $R$ is monotonically increasing. As we have just
seen, it would not be wise to assume this form for the string metric.
The coordinates would then break down outside the horizon at the
throat of the wormhole.  The $tt$ and $rr$ components of Einstein's
equation now become
\eqna\eteom
$$\eqalignno{   -rh' + 1-h & = r^2 h {\phi'}^2 + r^2 m^2 \phi^2
                 + {Q^2e^{-2\phi}\over r^2} \>,& \eteom a \cr
               {rhf'\over f} - 1+h & = r^2 h {\phi'}^2 - r^2 m^2 \phi^2
	        -{Q^2e^{-2\phi}\over r^2} \>.& \eteom b \cr} $$
These are both first order equations. The first is the standard
constraint equation and hence is independent of $f$. The remaining
component of Einstein's equation will be satisfied if the stress
tensor is conserved, which will hold if the dilaton satisfies its
field equation:
\eqn\ddeom{h^{1/2} [ h^{1/2} r^2 \phi']' + {r^2 h\phi' f'\over 2f}
	   = r^2 m^2 \phi -
	{Q^2e^{-2\phi}\over r^2}  \>.}
One can solve for $f$ by adding eqs.~\eteom{a} and \eteom{b}. The
result is
\eqn\fsol{ f = h e^{2\int r {\phi'}^2} \>.}
Thus $f$ vanishes if and only if $h$ does. We can now eliminate $f$
and $h'$ from the dilaton equation \ddeom\ to obtain:
\eqn\dtrue{ hr\phi'' +
          \phi'\[1+h - r^2 m^2 \phi^2 -{Q^2 e^{-2\phi}\over r^2}\]
	 -r m^2 \phi +{Q^2 e^{-2\phi}\over r^3} = 0 \>.}

The problem of finding black hole solutions thus reduces to solving
\eteom{a} and \dtrue. One can eliminate $m$ from these equations by
using the fact that they are invariant under the following rescaling:
$r\a cr, m \a m/c, Q\a cQ$. {\it We therefore set $m=1$.} This corresponds
to measuring distances in units of the Compton wavelength of the
dilaton.  Since the right hand side of \eteom{a} is positive definite,
it follows that $h$ cannot have an extremum with $h>1$.  If $h$ ever
increases  larger than one, it must continue to increase.

We now briefly consider the asymptotic form of the solutions near the
singularity. Translating the massless dilaton solution \massless\ near
the singularity into our new coordinates yields
\eqn\limit {h= a/r^2 \quad , \quad e^{-2\phi} = b \, r^2}
where $a$ and $b$ are constants related to $M$ and $Q$.  More
generally, let us try a solution of the form
\eqn\newlimit{ h = c/r^\alpha \quad , \quad e^{-2\phi} =
                                           d\, r^{2\beta} \>. }
where $c,d,\alpha$ and $\beta$ are constants.  The mass terms and charge
terms in
\eteom{a} and \dtrue\ will both be negligible near the singularity
if $\alpha$ is positive and larger than $ 2(1-\beta)$. In
this case, the leading order terms in \eteom{a} diverge like
$r^{-\alpha}$ and vanish if $\alpha = 1+\beta^2$. Combining this with
the condition $\alpha >  2(1-\beta)$ yields $\beta > \sqrt 2 -1$.
The leading order
terms in \dtrue\ diverge like $r^{-\alpha -1}$ and are automatically
satisfied. These solutions can be characterized by the constant
$\beta = - r\phi'$. By transforming the general massless solution \general\
near the singularity into our new coordinates, one can relate $\beta$ to
the constant $c_1$. The familiar massless
dilaton black hole has $\beta= 1$. We
will see that the massive dilaton black holes typically approach this
more general solution near the singularity with $\beta \ne 1$.
Since the radius of the spheres in the string metric is $r e^\phi
\sim r^{1-\beta}$, when $\beta >1$, the spheres become infinitely
large near the singularity.

We now need to numerically integrate the equations \eteom{a} and
\dtrue.  Since eq.~\eteom{a} is first order in $h$ and eq.~\dtrue\ is
second order in $\phi$, we must impose boundary conditions at a point
$r_0$ by specifying $h(r_0)$, $\phi(r_0)$ and $\phi'(r_0)$.  It might
seem straightforward to use the known asymptotic behavior of the
solutions to fix these three parameters at some large $r$ and then
integrate in. However this fails for the following reason. At large
$r$, the dilaton equation has both exponentially growing and
exponentially decaying solutions. Any small numerical error in the
initial values of $h(r_0)$, $\phi(r_0)$ and $\phi'(r_0)$ will
correspond to a small error in the coefficient of the mode which grows
exponentially as one evolves inward.  The result is that numerical
integration is extremely unstable and one cannot reliably integrate in
to the horizon. To avoid this, we use a technique which actually turns
the above instability to our advantage.\foot{We thank P. Bizon for
bringing this technique to our attention. It was used e.g. in \bizon.}

The first step is to reduce the number of free parameters.  A horizon,
where $h(r_0) = 0$, is a singular point of the equations \eteom{a} and
\dtrue.  Since $\phi''(r_0)$ appears multiplied by $h(r_0)$ in
eq.~\dtrue, at a horizon, eq.~\dtrue\ becomes a constraint relating
$\phi'(r_0)$ with $\phi(r_0)$. This reduces the number of parameters
needed to specify the solution by one. Since the horizon is a singular
point of the equations however, we cannot begin numerical integration
there.  Instead, we must do a power series expansion of the solution
around the horizon, and start the numerical integration slightly outside.
In other words, choose a point $r_0$. Expand the dilation
and metric around the horizon using
\eqn\expand{
\eqalign{ h(r) & = (r - r_0) \sum_{i=0}^{\infty} h_i (r - r_0)^i \>,\cr
       \phi(r) & = \sum_{i=0}^{\infty}  \phi_i (r - r_0)^i \>.\cr}}
Using this expansion in eq.~\eteom{a} and \dtrue, one sees that the
solution has one free parameter, $\phi(r_0) = \phi_0$. As happens with
most power series in this problem, the expansion \expand\ seems to be
only an asymptotic expansion, so in the numerical integrations, we
will keep only the first three terms. Even the low order terms are
quite lengthy, so we will not present them explicitly here.

The numerical integration now proceeds as follows. Fix $Q$, and select
a value for $r_0$. This fixes the area of the horizon which is
equivalent to fixing a mass $M$. (Recall that we have set $m=1$.)
Choose a small parameter $\epsilon$ which determines the starting
point of the integration $r_0 + \epsilon$. Now choose a value of
$\phi_0$ (this determines $h(r_0 + \epsilon)$, $\phi(r_0 + \epsilon)$,
and $\phi'(r_0+\epsilon)$) and integrate outwards. As we have seen,
the equations \eteom{a} and \dtrue\ are extremely numerically
unstable, which makes the outwards integration tricky.  One can use
this instability to determine the correct value for $\phi_0$.  The
homogeneous solution to the flat space dilaton equation \dilasym\ has
an exponentially growing solution. If $\phi_0$ is chosen to be too
small, then the undesired exponential mode mixes into the real
solution with a negative coefficient. As $r$ increases, this drives
$\phi(r)$ off towards $-\infty$.  As discussed in Sec.~2, $\phi(r)$
cannot be negative outside an event horizon. Thus it is easy to tell
if the initial value of $\phi_0$ was too small. Conversely, if
$\phi_0$ is initially chosen to be too large, then the undesired
exponential mode enters with a positive coefficient, which drives
$\phi(r)$ towards $+\infty$. As we saw in Sec.~2, $\phi(r)$ must
monotonically decrease outside the horizon, so it is also easy to tell
if one has chosen $\phi_0$ too large. Once one has found a $\phi_0$
that is too large and one that is too small, it is straightforward to
test their mean value to see whether it is too large or too small.
One can then iterate the procedure until a value of $\phi_0$ is found
that approximates the desired asymptotic behavior. Because of
numerical errors, it is impossible to completely eliminate the
instability, but it is possible to determine the correct value of
$\phi_0$ to extremely high precision.

To obtain the interior solution, one uses the value of $\phi_0$ just
obtained and integrates inward starting at $r_0 - \epsilon$.  The
crucial point is that $\phi_0$ can be determined to sufficient
accuracy so that the interior solution is insensitive to the residual
error.  For small black holes, one can integrate the equations to
arbitrarily close to the singularity. For black holes with an inner
horizon, the equations become singular at the inner horizon, and the
integration breaks down there. However, one can integrate to quite
close to the inner horizon, and determine the value of $\phi$ at the
inner horizon.  This can be used as a new $\phi_0$ to integrate
inwards to the singularity.  One must be careful with this procedure,
because the value of $\phi$ at the inner horizon cannot be determined
as precisely as the value of $\phi$ on the event horizon, and the
solutions near the singularity are sensitive to the value of
$\phi_0$ used at the inner horizon.

To perform the numerical integration, we used the extremely powerful
differential equation solver {\it lsoda.f}\ \hindmarsh.  Using this
Fortran program, it takes very little time to determine the correct
value of $\phi_0$ to up to fourteen decimal places in all but the most
pathological situations.

\subsec{Specific Numerical Examples}

\ifig\bigdiv{Integrating the dilaton outside an event horizon
with $r_0 = 100$ and $Q = 60$. The upper line is
$\phi_0 = 3.5988199312 \times 10^{-5}$, the middle (dashed) line
is ${Q^2 / r^4}$, and the bottom line is
$\phi_0 = 3.5988199311 \times 10^{-5}$.}{\epsfbox{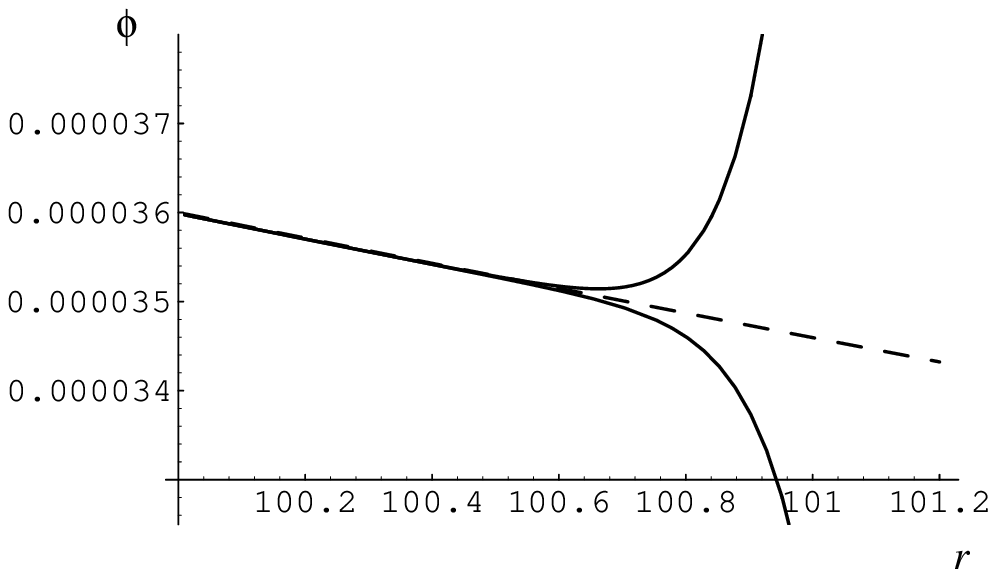}}

For our first example, we take $r_0 = 100$ and $Q = 60$. This corresponds
to a mass
$M=68$, which can be determined from the fact that outside the
horizon, this solution is very close to the \RN\ solution. These parameters
should
correspond to a large black hole with two horizons.  In~\bigdiv,
we show what happens when one integrates outwards.  Two integrations
are shown, with slightly different values of $\phi_0$.  One is
diverging upwards, and the other is diverging downwards. The dashed
line in the middle is just the approximation from eq.~\dilbsym,
showing that the dilaton is quite close to the expected value outside
the event horizon (the leading order terms in~\dilbsym\ and \asym{a}
are still correct in this coordinate system).  The correct value for
$\phi_0$ lies between the two given values.  Notice that the outwards
integration does not go far before the instability sets in. This is
not a serious problem, since we have a good approximation to the
solution outside the horizon,~\dilbsym\ and~\asym{}.  The important
point is that one can integrate far enough to determine $\phi_0$ to
high precision.  Integrating inwards from the event horizon works
well, and the two values of $\phi_0$ give essentially
indistinguishable answers inside the black hole, at least up to the
inner horizon.

\ifig\bigh{$h(r)$ between the horizons of a black hole with
$r_0 = 100$ and $Q = 60$.}{\epsfbox{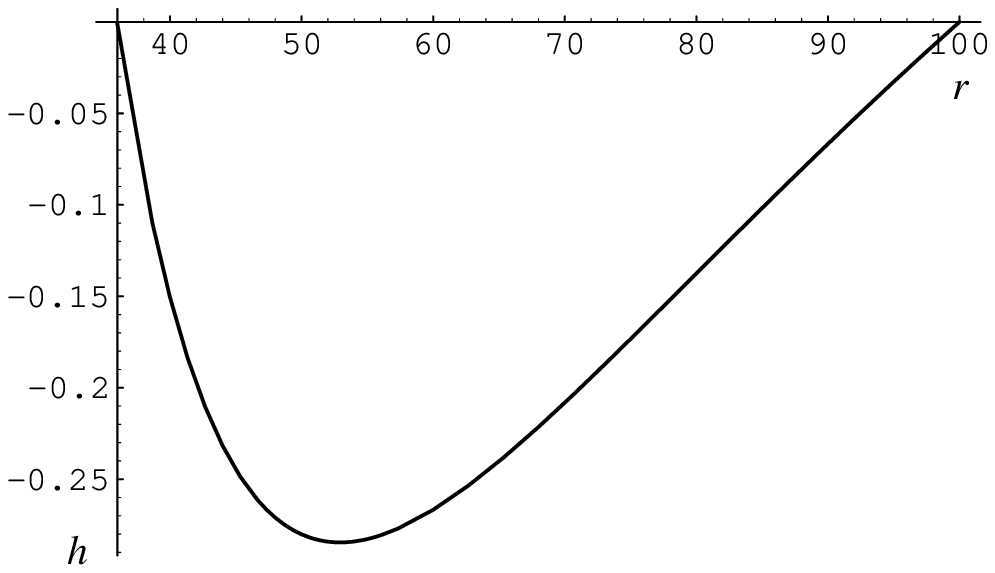}}

\ifig\bigd{$\phi(r)$ between the horizons of a black hole with
$r_0 = 100$ and $Q = 60$.}{\epsfbox{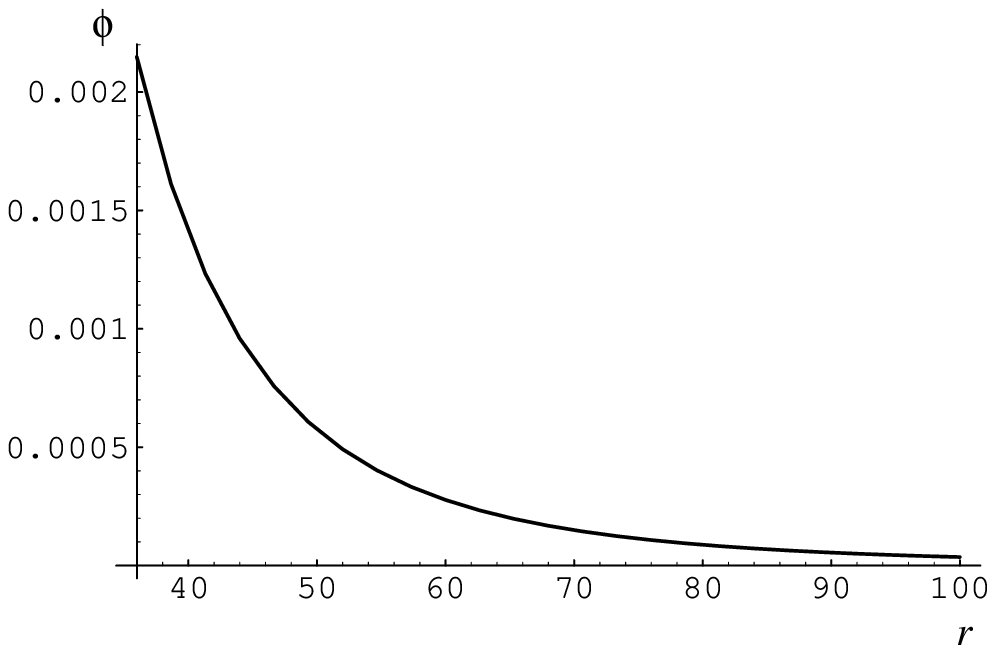}}

The metric function $h(r)$ for this black hole inside the event horizon
is shown in~\bigh.  It
agrees almost precisely with the approximate solution in~\asym{a}.
The deviations are too small to show up on the plot.  The inner
horizon appears at $r = 35.976$. This confirms that black holes with
massive dilatons can have inner horizons. The dilaton is shown in~\bigd. It
also has essentially no deviation from the approximate value~\dilbsym.

\ifig\bigoops{$\phi(r)$ inside the inner horizon. The upper line
is the numerical result with $\phi_0 = 2.1517496625 \times 10^{-3}$,
the lower line is $\phi_0 = 2.1517496620 \times 10^{-3}$,
and the dashed line is the expected approximate
value from~\dilbsym.}{\epsfbox{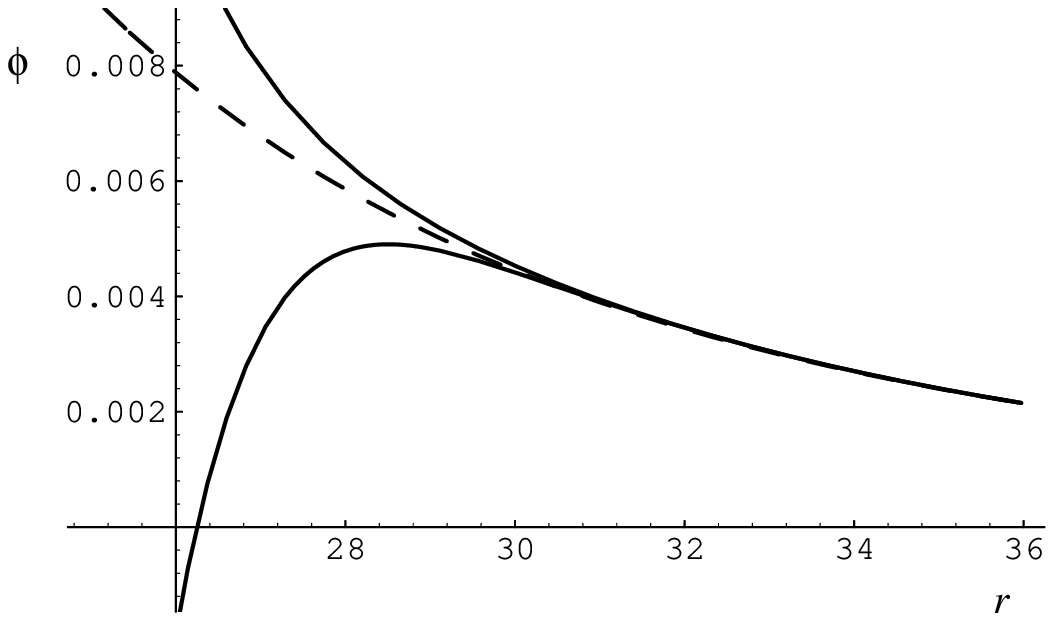}}

Continuing the integration inside the inner horizon is more difficult.
Because the equations become singular at the inner horizon, it is
simple to determine its position. Unfortunately, numerical errors make
the precise value of $\phi$ difficult to determine.  A good
approximation can be found, but this seems to be insufficient to
integrate up to the singularity, as the interior solution exhibits an
instability similar to the one outside the event horizon.  (This
instability does not occur when the inner horizon is absent, and seems
to be related to $h>0$.)  An example of this is shown in~\bigoops,
where two similar values of $\phi$ on the inner horizon are used to
begin integration. In most cases, one can push the instability to
inside the point where $h(r) > 1$. We have seen that once $h(r) > 1$,
it must increase to $+\infty$, so there can be no third horizon.

\ifig\bignoh{$h(r)$
for a black hole with $r_0 = 100$ and $Q = 20$,
which corresponds to $M = 52$. The solid line is the numerical
result, and the dashed line is the approximation from~\asym{a}.
}{\epsfbox{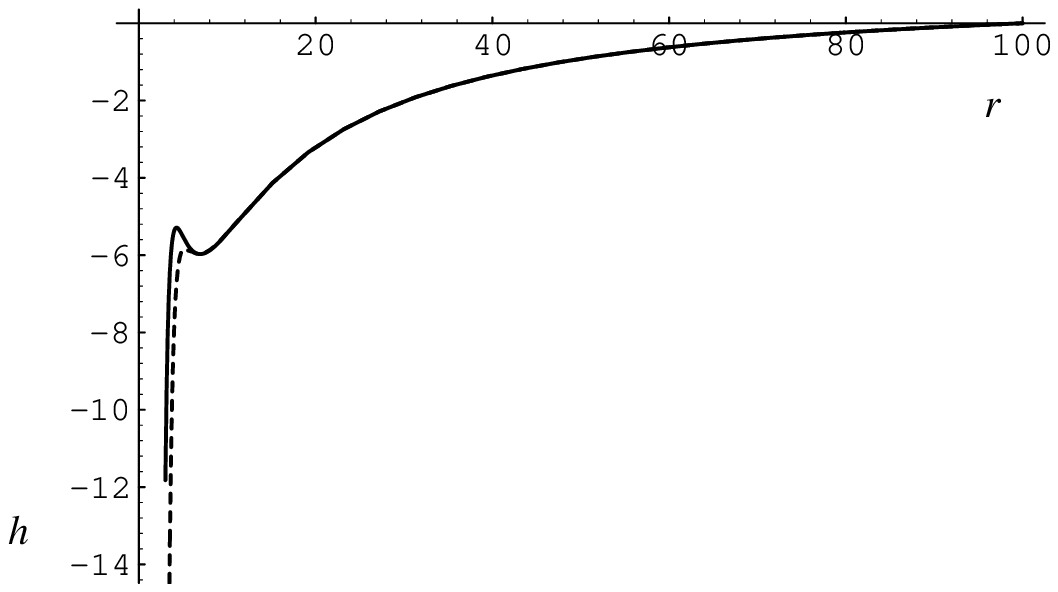}}

Our next example is a large black hole which has no inner horizon.
The event horizon is again chosen to be at $r_0 = 100$, but the charge
is now $Q = 20$.
These parameters correspond to a black hole with mass $M =
52.$ Again, it is not possible to integrate outwards very far from the
event horizon, but $\phi_0$ can be determined to high accuracy.
We find that with these parameters, the dilaton at the
horizon is $\phi_0 = 3.998433496484 \times 10^{-6}$, which is
excellent agreement with the approximate answer given by $Q^2/ r^4 = 4
\times 10^{-6}$.
Unlike the previous example, the solution inside the event horizon is
quite stable under small changes in $\phi_0$ all the way to the
singularity.  In~\bignoh, we compare the metric function $h(r)$ inside
the horizon with the approximate solution given in
equation~\asym{a}. The agreement is excellent in to $r \ap 5$. The
small increase in $h(r)$ around $r=4$ is where the \RN\ solution has a
inner horizon.

\ifig\bignod{$\log \phi(r)$ for a black hole with $r_0 = 100$ and $Q = 20$.
The solid line is the numerical solution, and the dashed line
is the approximate solution from~\dilbsym.}{\epsfbox{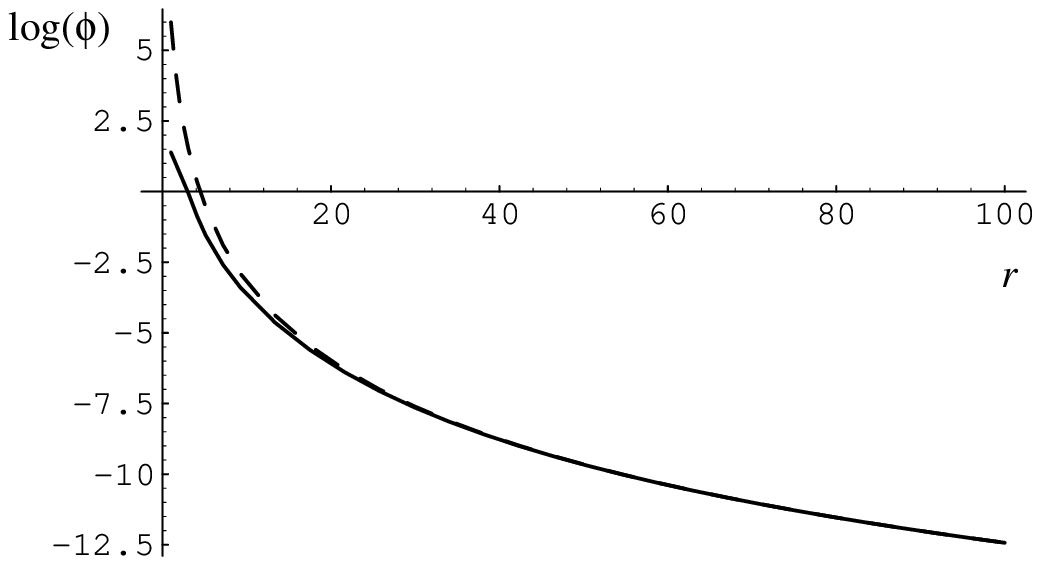}}

In~\bignod, we show the dilaton inside the event horizon of the same
black hole. Since the dilaton varies across a wide range, we have
actually plotted $\log \phi$. For comparison, we have also plotted
$Q^2 / r^4$. The dilaton agrees  with the approximate solution
until well inside the event horizon, but as expected, starts diverging
inside $r \ap 10$ (this deviation is {\it not} numerical error, but due to
the breakdown of the approximation~\dilbsym).

\ifig\bignodp{The behavior of $r \phi'(r)$ near the singularity when
$r_0 = 100$ and $Q = 20$.}
{\epsfbox{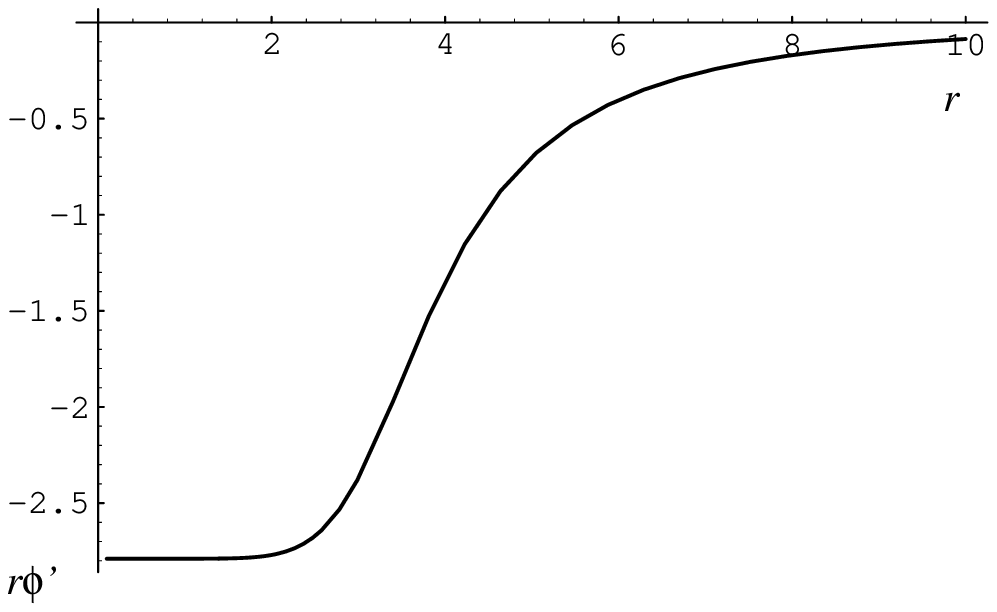}}

The behavior of this black hole near the singularity is quite
illustrative.  The behavior of $r \phi'(r)$ is shown in~\bignodp. As
discussed earlier, if the solution were approaching the massless
solution~\massless, near the singularity, $r \phi' \rightarrow -1$. We
can see clearly that this is not the case. Instead, $r \phi'
\rightarrow -2.79$.  This corresponds to a solution of the
type~\newlimit, so the mass terms are becoming negligible near the
singularity. (As expected, $h$ diverges like $r^{-(1+\beta^2)}$ with
$\beta = 2.79$.) However, the solution is {\it not} approaching the
massless solution~\massless.  Instead, it is approaching a more general
solution to
the massless equations.

\ifig\smallho{$h(r)$ outside the event horizon of a small black hole
with $r_0 = .01$ and $Q = .001$.}{\epsfbox{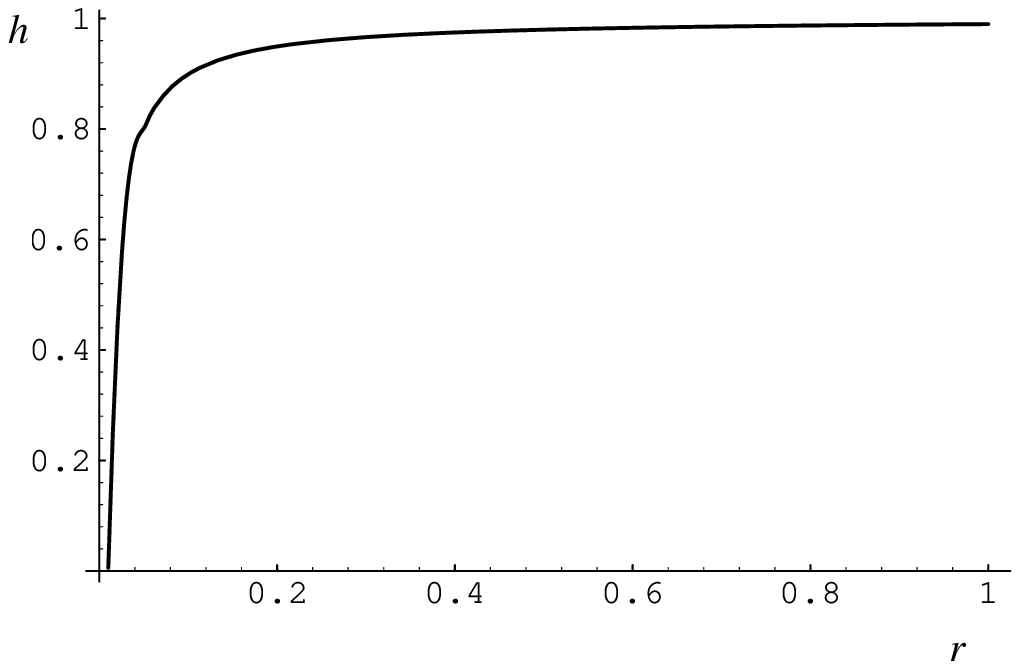}}

The metric function $h(r)$ for a small black hole is shown
in~\smallho.  This black hole has $r_0 = .01$ and $Q = .001$. For small
black holes, one can integrate outwards considerably farther before
the numerical instability sets in.
The
mass $M$ of the black hole can be easily determined to be $M =
.00504976$ by integrating out to  a large enough $r$ so that
the asymptotic approximation~\asym{a} is valid.

\ifig\smalldo{$\log \phi(r)$ outside the event horizon of a black hole
with $r_0 = .01$  and $Q = .001$. The solid line is the
numerical result, the upper dashed line is the massless solution,
and the lower dashed line is the asymptotic solution.}{\epsfbox{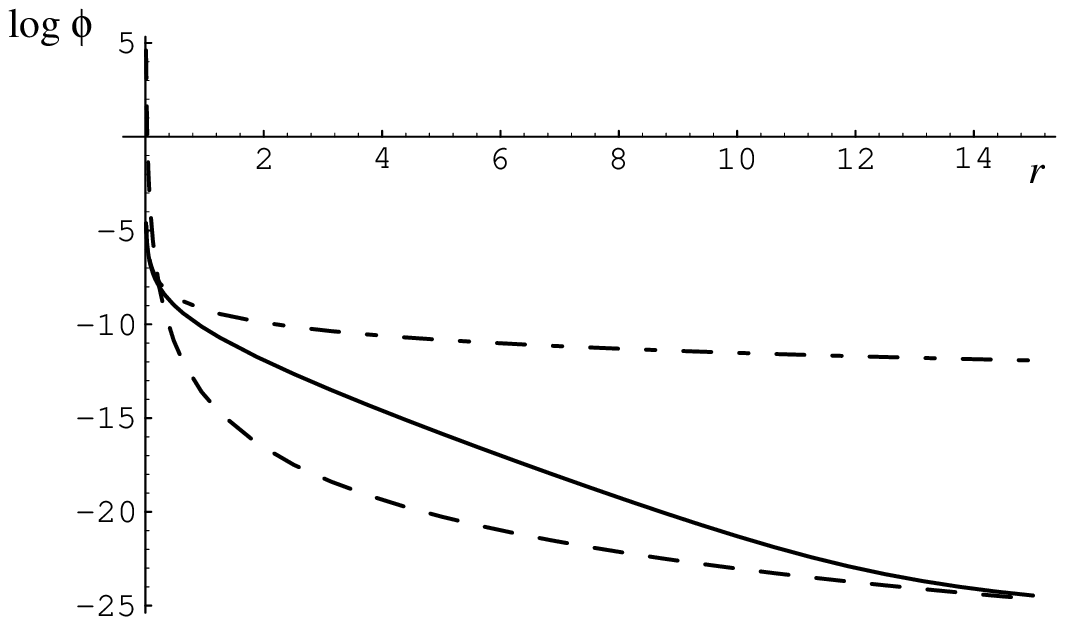}}

In~\smalldo, we show the dilaton outside the event horizon for the
same small black hole, along with the asymptotic solution~\dilbsym\
and the massless solution~\massless\ for this $Q$
and $M$ (setting $\phi_{\infty} = 0$ to match properly at the horizon, and
transforming into the new coordinates~\einthree).
As expected, the dilaton is
falling off much faster than in the massless dilaton solution. At large
$r$ it  clearly
approaches the asymptotic solution.

\ifig\smalldi{$\log \phi$ inside the event horizon of a small black
hole with $r_0 = .01$ and $Q = .001$.}{\epsfbox{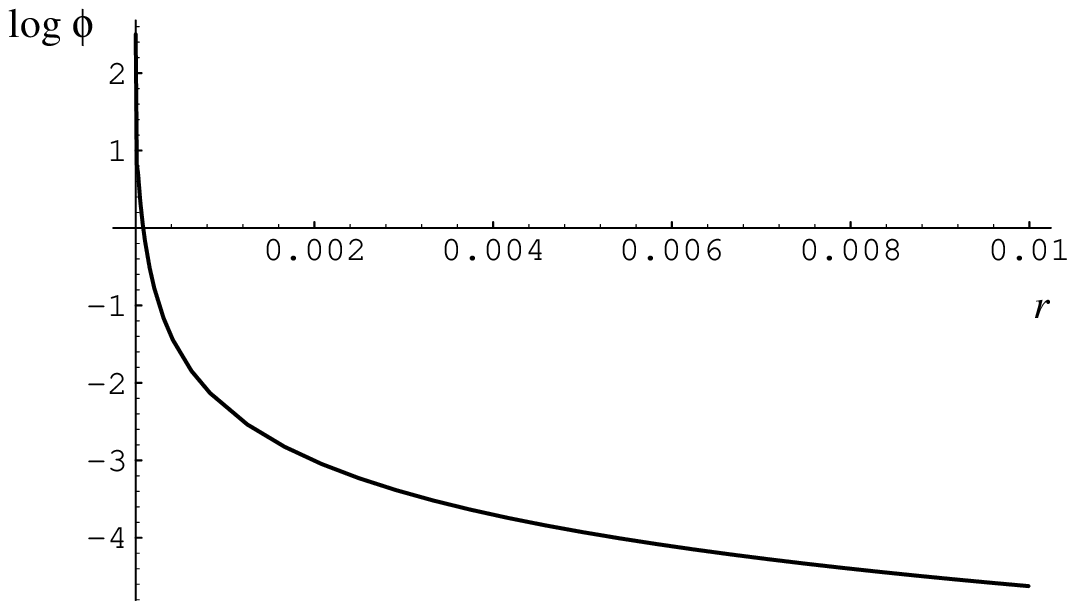}}

Inside the event horizon of this small black hole, the situation is different.
Now the numerical solution
behaves almost precisely like the solution with a massless
dilaton. The
dilaton inside is shown in~\smalldi. The dilaton for the $m=0$ solution
is indistinguishable in this plot. This confirms that small black holes
behave similarly to the $m=0$ solutions. In this case, one finds $r \phi'
\rightarrow -1.00000097$ at the singularity,
which is close to the $m=0$ value. The
interior solution is stable under perturbations of $\phi_0$, so the
slight deviation from $r \phi' \rightarrow -1$ does seem to be real.
Since this black hole is not extremal, some deviation might be expected.
The fact that it is so small seems to be a reflection of the fact that
$Q$ is quite small.

\ifig\medha{$h(r)$ for a black hole with $r_0 = .85$ and $Q = 1.36455$.}
{\epsfbox{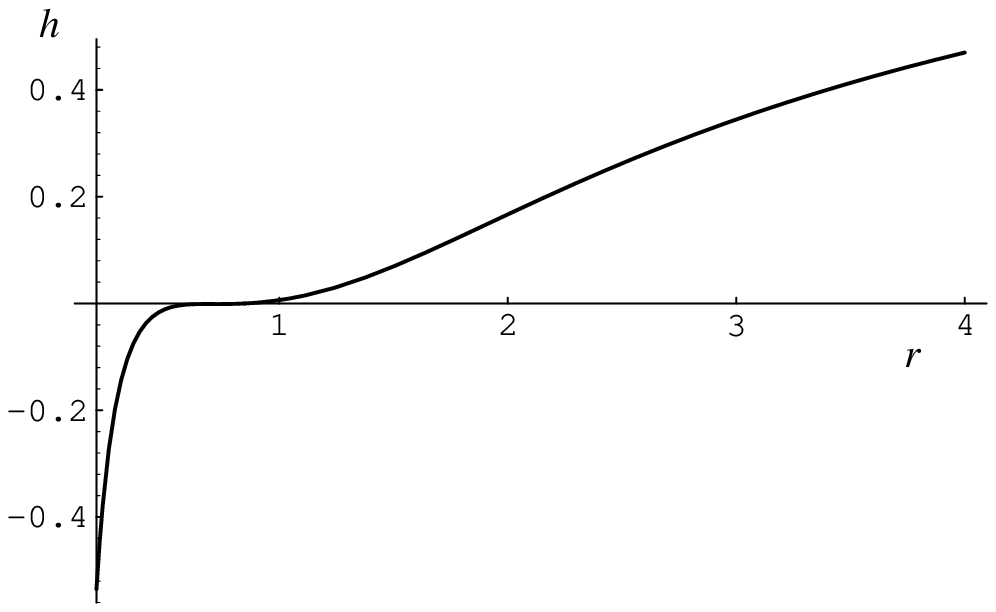}}

\ifig\medhb{An enlargement of $h(r)$ near the horizon for the medium black
hole.}{\epsfbox{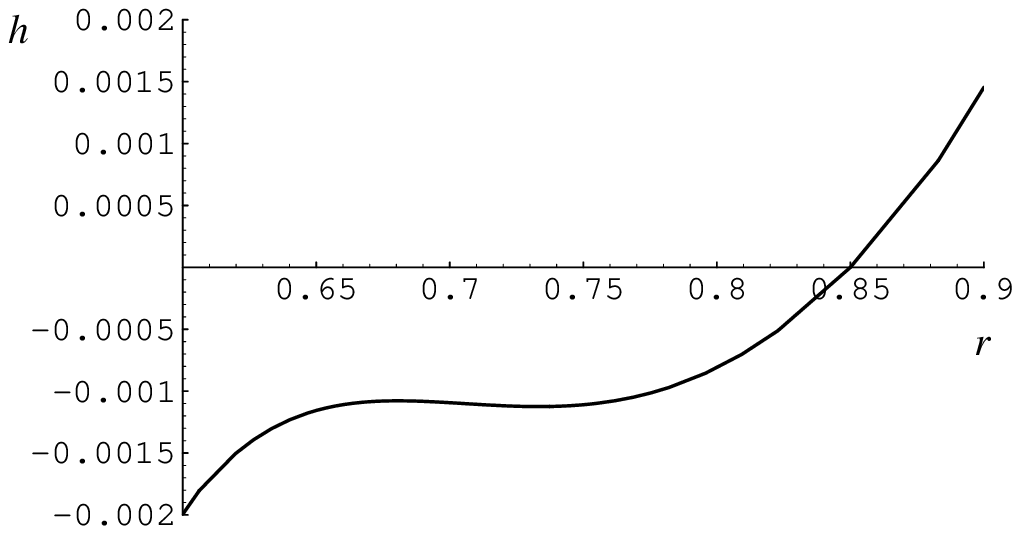}}

Our next example is a medium sized black hole, with $r_0 = .85$ and $Q
=1.36455$. The metric function near the horizon is shown in \medha,
and a more detailed picture is in \medhb. From the asymptotic value of
$h(r)$, we can determine the mass of this black hole to be $M =
1.29283$. This black hole is close to the extremal limit, and also
close to the triple point. Notice that $Q^2/M^2 > 1$, so this black
hole is unstable to splitting into smaller, widely separated black
holes.  If we were to decrease $r_0$ or increase $Q$ by a small
amount, the slight increase in $h(r)$ would touch $h=0$, causing an
inner horizon.

\ifig\medworm{The radius of spheres in the string metric, $r e^{\phi}$,
is shown near the horizon for the medium black hole. The straight line
is the position of the horizon.}{\epsfbox{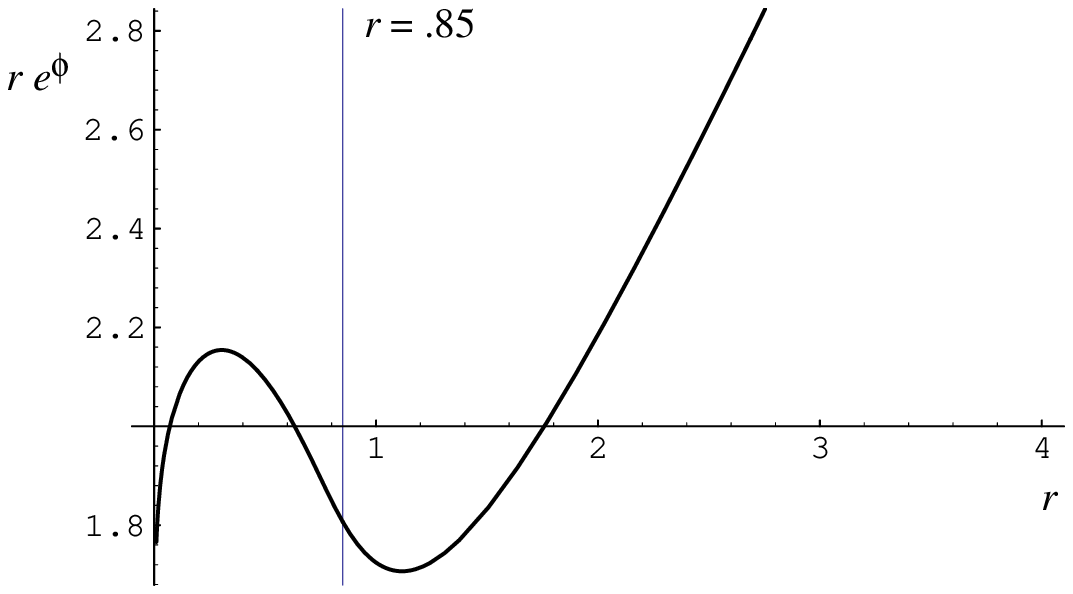}}

The horizon for this black hole lies behind a wormhole in the string
metric. In \medworm, we show the radius of spheres in the string
metric $r e^{\phi}$ as a function of $r$. Notice that $r e^{\phi}$
reaches a minimum {\it outside} of the event horizon, and increases
inwards until turning over again farther in.

\ifig\extapp{$\beta$ {\it versus} $\log r_0$, the position of the
event horizon, for black holes with charge $Q = 1$.}{\epsfbox{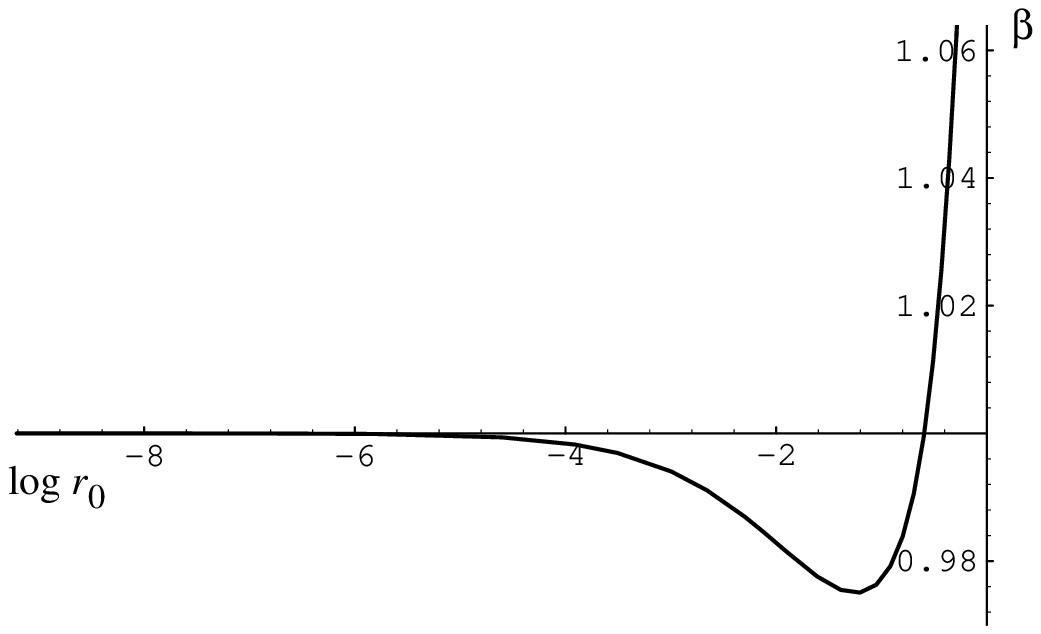}}

We demonstrate the approach to the extremal limit in \extapp.  This
shows the results for a number of black holes, all with charge $Q =
1$.  The horizon for a black hole is picked to be at a particular
$r_0$, and the value of $\beta = -\lim_{r \rightarrow 0} r \phi'$
is then determined.  Smaller values of $r_0$ correspond to black holes
closer to the extremal limit. As can be seen in \extapp, as the black
hole approaches the extremal limit, $\beta \rightarrow 1$, which is the
massless dilaton value.
Thus, even medium
black holes with a charge too small to have a degenerate horizon, will have
an infinite throat in the extremal limit. Notice that $\beta$ is becoming
larger than one as one moves
farther from the extremal limit. This is consistent
with our earlier result that $\beta = 2.79$ for a black hole with $Q=20$.
As the extremal limit is approached, we can
determine from $h(r)$ that $M \rightarrow .8928$. This corresponds to
$Q^2/M^2 = 1.2545$ which is in agreement with the  statement that this ratio
varies from one for large black holes to two for small holes.

\ifig\notext{The radius of spheres in the string metric as a function of
the proper distance $\rho$
for a black hole with $r_0 = 10^{-2}$
and $Q = 10^{-2}$. The vertical line is the horizon at
$\rho = 1.08102$.}{\epsfbox{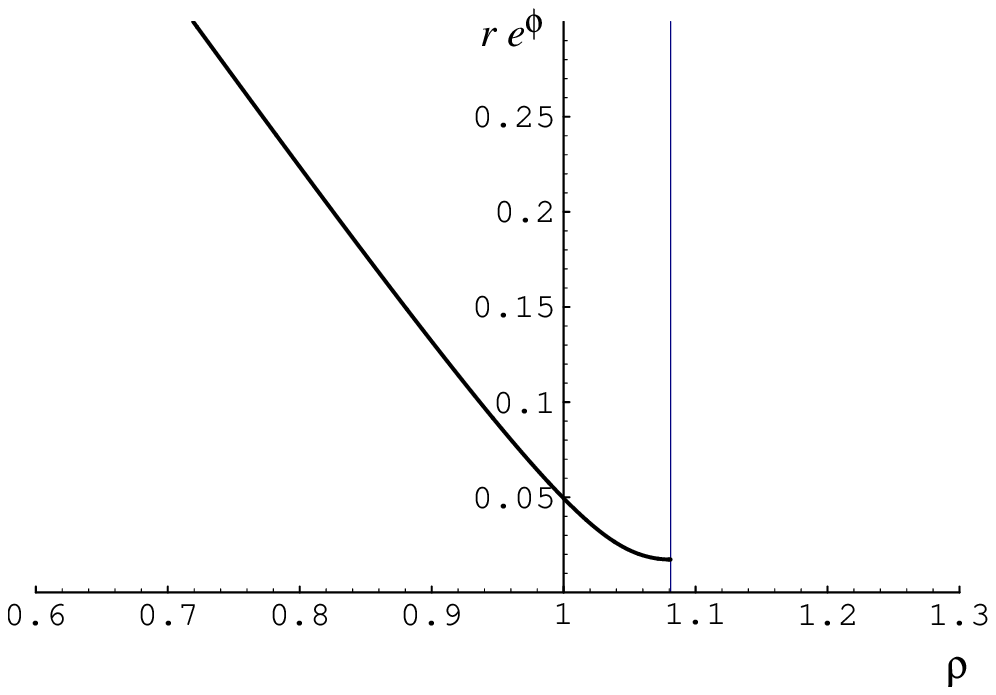}}

\ifig\nearext{The radius of spheres in the string metric as a function
of the proper distance $\rho$
for a black hole with $r_0 = 10^{-5}$
and $Q = 10^{-2}$. The vertical line is the horizon at
$\rho = 1.27191$.}{\epsfbox{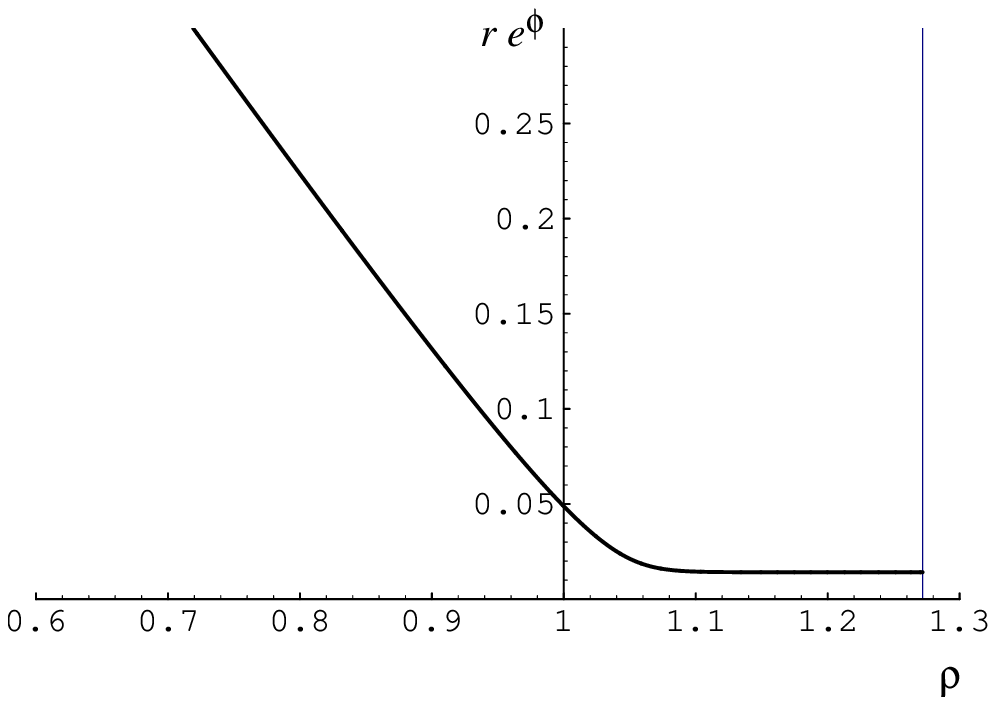}}

We conclude this section by demonstrating the formation of a long
throat outside the event horizon of a nearly extremal black hole.  The
geodesic distance $\rho$ from a point $r_1$ to a point $r_2$ in the
string metric is
\eqn\geod{ \rho = \int_{r_1}^{r_2} {e^{\phi} \over \sqrt{h}} dr \>.}
We will arbitrarily choose $r_2 = 1$, and show in \notext\ and
\nearext\ the radius of the spheres in the string metric $r e^{\phi}$
as a function of the geodesic distance $\rho$ for two different
black holes.
The black hole in
\notext\ has $r_0 = 10^{-2}$ and $Q = 10^{-2}$, so it is not yet near
its extremal limit. The mass of this solution is $M = .0086768$.
Notice that $r e^\phi$ is basically linear until just before the
horizon at $\rho = 1.08102$. Since $\rho$ measures proper distance inward from
a fixed  radius outside the horizon,
this plot is reversed compared to some of the
earlier ones. Increasing $\rho$ corresponds to decreasing radius.

In \nearext, we show a black hole much
closer to its extremal limit, with $r_0 = 10^{-5}$ and $Q = 10^{-2}$,
which correspond to a mass of $M = .007096$. In terms of proper
distance, the horizon is at $\rho = 1.27191$. For this black hole, $r
e^\phi$ levels off before the event horizon.  Thus the horizon is down
a throat of essentially constant radius.  The length of the throat grows
as the black hole approaches its extremal limit.  Note that the plots
\notext\ and \nearext\ are virtually identical except that the horizon
is moved farther down the throat.

\newsec{Conclusions}

We have studied the behavior of black holes coupled to a massive
dilaton.  Although exact solutions are not known in closed form, we
have obtained a fairly complete picture of their properties by
combining analytic arguments with numerical integration. One of our
main results is that even with a mass, the dilaton can have an
important effect in the late stages of black hole evaporation. For
black holes with\foot{In this expression, $e$ is the base of the natural
logarithm
 and not the charge on the electron!}
$0<|Qm|< e/2$, the spacetime initially (when the mass
is large) is close to the \RN\ solution. As the mass decreases, the
spacetime becomes closer to the massless dilaton solution. So all of the
unusual features of the extremal $m=0$ black holes should emerge in the
late stages of the evaporation. The
transition occurs when the curvature is still small compared to the
Planck scale so higher order string corrections should be negligible.
If $|Qm|> e/2$ and $M \gg Q$, then the black hole solutions have only
one horizon. But as the mass decreases, there comes a point where an inner
horizon appears. Further evaporation will then be similar to \RN\
with the extremal limit containing a double horizon.

We have considered only the simplest potential for the dilaton $m^2
\phi^2$.  Even though the dilaton is becoming large, this potential
becomes negligible near small black holes since other terms in the
field equations diverge more quickly. This shows that any potential
which grows more slowly than $m^2 \phi^2$ for large $\phi$ should have
black hole solutions with similar behavior. (The limit on the
charge $|Qm|> e/2$  for the existence of a double horizon will presumably
be modified.) Although the exact
dilaton potential in string theory is not known, there have been
proposals based on gluino condensation \dine.  These potentials
typically approach a constant for large $|\phi|$ and so the black holes
should be similar to the case we have analyzed. Also, these potentials
are typically not symmetric under $\phi \a -\phi$. This implies that
electric and magnetic charged black holes will no longer be related by
a duality symmetry.

We have shown that nearly extremal black holes coupled to a dilaton
are repulsive. Roughly speaking, this is because the presence of the
dilaton near the horizon
allows $Q^2 >M^2$, but the dilaton mass cuts off the attractive dilaton force
at large distances.
This result depends only on properties
of the dilaton potential near its minimum, and will hold for any
potential that gives the dilaton a mass.
We  also argued that one consequence of this is
that large extremal black holes are quantum
mechanically unstable to bifurcation.
Since charge is quantized,
the extremal black hole of smallest charge must be stable. It will
repel black holes with the same charge and attract those with opposite
charge. It is interesting to compare this to an elementary particle.
The standard argument that the electron cannot be a black hole is that
its charge is much larger than its mass. So according to the \RN\
solution, it would have to be a naked singularity. However, we have seen
that this is
not the case when massive dilatons are taken into account.  (Other
similarities between black holes with dilatons and elementary
particles have recently been discussed in \wilczek.)  Could all
elementary particles be black holes?  One obvious objection is that we
have found that black holes can have their charge slightly larger than
their mass, while the ratio of the charge to mass of an electron is
about $10^{21}$. However we have only considered the case where the
dilaton vanishes at infinity.  For the magnetically charged black
holes that we have been discussing, if $\phi \a \phi_{\infty}$
asymptotically, the maximal charge to mass ratio is increased by
$e^{\phi_{\infty}}$. For electrically charged black holes, the sign of
$\phi$ is reversed. So if $\phi_{\infty}$ is negative at infinity,
corresponding to weak coupling, the charge to mass ratio can be much
larger than one.  Another obvious objection is that elementary
particles with the same charge can have different masses. Perhaps this
could be accounted for when the other interactions are included.  Even
if the known elementary particles are not small black holes, it is
certainly intriguing that in many respects, they act qualitatively the
same.

We have also shown that the string metric describing black holes with
$Mm \ap 1$ can have wormholes outside the event horizon. It is not
clear how sensitive these are to the details of the dilaton potential.
However it should be kept in mind that we have discussed only
classical solutions.  These wormholes only occur when $\phi \ap 1$,
which is where string loop corrections may become important.

\bigskip\centerline{Note added}

As this paper was being completed, we received a preprint from R.
Gregory and J. Harvey, ``Black Holes with a Massive Dilaton,''
hep-th/9209070, which overlaps with some of this work.

\bigskip\centerline{Acknowledgements}

It is a pleasure to thank P. Bizon, S. Giddings, R. Gregory, J.
Harvey, and A. Strominger for discussions.  We also wish to thank the
Aspen Center for Physics.  This work was supported in part by NSF
Grant PHY-9008502 and by DOE grant DE-AC02-76ER03075.

\listrefs

\bye